\documentclass[twoside]{ilcws08}
\usepackage[latin1]{inputenc}
\usepackage[dvips]{graphicx,epsfig,color}
\usepackage{wrapfig,rotating}
\usepackage{amssymb,amsmath,array}

\begin{document}
\title{
Luminosity Performance Studies of Linear Colliders with Intra-train Feedback Systems} 
\author{J.~Resta-L\'opez$^1$, P.~N.~Burrows$^1$, A.~Latina$^2$ and D.~Schulte$^3$
\vspace{.3cm}\\
1- John Adams Institute at Oxford University, Oxford, OX1 3RH, UK
\vspace{.1cm}\\
2- Fermilab, Batavia, IL 60510-5011, USA
\vspace{.1cm}\\
3- CERN, 1211 Geneva 23, CH
}

\maketitle

\begin{abstract}
The design luminosity for the future linear colliders is very demanding and challenging. Beam-based feedback systems will be required to achieve the necessary beam-beam stability and steer the two beams into collision. In particular we have studied the luminosity performance improvement by intra-train beam-based feedback systems for position and angle corrections at the interaction point. We have set up a simulation model which introduces different machine imperfections and can be applied to both the International Linear Collider (ILC) and the Compact Linear Collider (CLIC). 

\end{abstract}

\section{Introduction}

\begin{wrapfigure}{r}{0.5\columnwidth}
\centerline{\includegraphics[width=0.3\columnwidth, angle=-90]{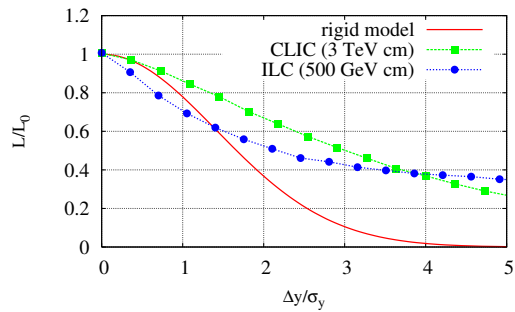}}
\vspace{-0.2cm}
\caption{Luminosity loss versus the relative beam-beam position offset.}\label{bbdeflection}
\end{wrapfigure}

The design luminosity of the future high-energy electron-positron colliders requires transverse beam sizes at the nanometre level at the interaction point (IP), as well as stabilisation of the electron and positron beam collision overlaps at the sub-nanometre level. 

Static and dynamic imperfections of the machine generate beam jitter, which can dramatically degrade the luminosity. Fig.~\ref{bbdeflection} shows an example of luminosity loss versus the relative offset of the colliding beams for both ILC \cite{ILC} and CLIC \cite{CLIC} with nominal beam parameters. The red curve is the geometric luminosity analytically calculated. The blue and green curves correspond to the luminosity loss for ILC and CLIC, respectively, computed using the code GUINEA-PIG \cite{guinea}. For instance, in order to maintain the luminosity 10\% of nominal the electron and positron beam vertical position overlap needs to be stabilised to within $\approx 0.5~\sigma_y$ for ILC and $\approx 1~\sigma_y$ for CLIC.              

In order to achieve the required beam stability goals, beam-based feedback (FB) systems, operating at different timescales, are designed to be distributed in the linac and the beam delivery system (BDS). In this paper we concretely study the luminosity performance with fast intra-train beam-based FB systems at the IP in presence of dynamic imperfections, e.g. ground motion (GM). Simulation results are presented and discussed in section~\ref{simula} for both ILC and CLIC. 


\begin{table}
\footnotesize{\centerline{\begin{tabular}{|l|c|c|l|}
\hline
\textbf{Property}  & \textbf{ILC 500 GeV} & \textbf{CLIC 3 TeV} & units \\
\hline
Electrons/bunch & 2.0  & 0.37 & $10^{10}$\\
Bunches/train & 2820 & 312 & \\
Train Repetition Rate & 5 & 50 & Hz \\
Bunch Separation & 308 & 0.5 & ns \\
Train Length & 867.7 & 0.156 & $\mu$s \\
Horizontal IP Beam Size ($\sigma_{x}$) & 655 & 45 & nm \\
Vertical IP Beam Size ($\sigma_{y}$) &  5.7 & 0.9 &  nm \\
Longitudinal IP Beam Size &  300 & 45  & $\mu$m \\
Luminosity & 2.03 & 6.0 & $10^{34}$cm$^{-2}$s$^{-1}$ \\
\hline
\end{tabular}}}
\caption{Some nominal parameters of the ILC and CLIC designs for the linear collider.}
\label{tableparams}
\end{table}

To illustrate the train structure of linear colliders Table~\ref{tableparams} shows some relevant design parameters of ILC and CLIC. The ILC is an example of 'cold-'RF based design (superconducting RF cavities) \cite{ILC}. The ILC repetition frequency is 5 Hz, and the bunch train comprises 2820 bunches with a nominal inter-bunch separation of 308~ns. This time structure would allow the bunch-to-bunch feedback correction using digital FB processors. CLIC is based in a so-called 'warm'-RF design (see e.g. Ref.~\cite{CLIC} for details), with beam time structures much shorter than for the ILC. For CLIC with a nominal inter-bunch separation of 0.5~ns and a nominal train length of 156~ns the design of an IP intra-pulse FB system is very challenging.



\section{Intra-train feedback system}

\begin{wrapfigure}{r}{0.5\columnwidth}
\centerline{\includegraphics[width=0.45\columnwidth]{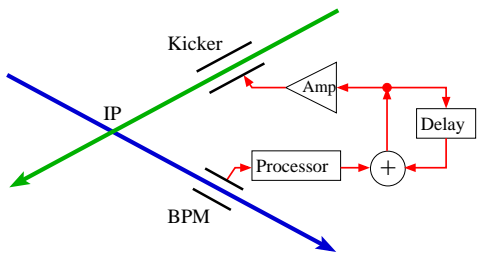}}
\caption{Schematic of IP intra-train feedback system.}\label{basicschematic}
\end{wrapfigure}

The beam-beam jitter induced by imperfections of the machine can be counteracted by using fast intra-train feedback systems near the IP. The IP-FB systems could be considered as the last 'weapon' of defence against, for example, transverse jitters of the final focus magnets. 

A fast IP-FB system for linear colliders is based on the measurement of the incoming trajectories of the early bunches in the electron or positron trains. This information is then used as the input to the feedback system for steering the later bunches into collisions at the IP. Fig.~\ref{basicschematic} shows a basic scheme of a beam-based IP-FB system.  The key components of the system are a beam position monitor (BPM) for registering the beam orbit, and a kicker for applying a position (or angle) correction to the beam. Additional hardware components, such as BPM signal processor board, FB circuit and fast amplifier, are described in detail elsewhere \cite{FONT}.

\section{Beam transport and luminosity simulations}
\label{simula}

We have developed a beam transport model for linear colliders based on the tracking code PLACET \cite{PLACET}. Sliced bunches are tracked along the main linac. The linac simulations include both short and long-range transverse and longitudinal wakefield effects of the accelerating cavities. For the tracking through the BDS each bunch is binned in a certain number of macro-particles (here we have used 50000 macro-particles). Survey alignment errors and dynamic imperfections (ground motion) can be included in the model. 



The code GUINEA-PIG \cite{guinea} is used to evaluate the luminosity and the beam-beam deflection for electron-positron beam collisions at the IP.    

\subsection{Results for ILC}

\begin{wrapfigure}{r}{0.5\columnwidth}
\centerline{\includegraphics[width=0.5\columnwidth]{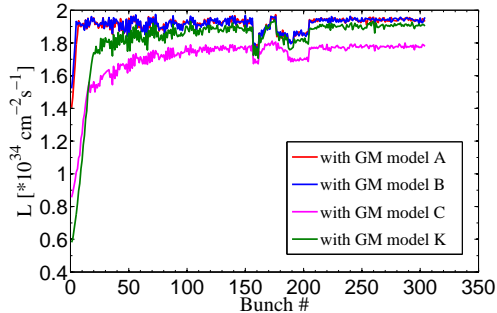}}
\vspace{-0.2cm}
\caption{Total luminosity versus bunch number within a train for the ILC. Example for a single random seed for different models of ground motion: A, B, C and K.}\label{ILCLuminosity}
\end{wrapfigure}

Intra-train IP position FB corrections can be applied using a stripline kicker located near the IP in the incoming beamline between the sextupole SD0 and the final quadrupole QF1. At $\pi/2$ phase advance downstream of the IP a BPM measures the beam positions to determine the deflection angle of the beams. Usually BPM resolutions of $\sim 1~\mu$m are sufficient for these measurements. On the other hand, to correct the angle, a stripline kicker is located at the entrance of the final focus system with a downstream BPM at $\pi/2$ phase advance. For the simulations we use a FB loop based in a proportional and integral (PI) control algorithm \cite{THimmel}. 

We have studied the performance of the FB system in terms of correcting the position jitter generated by ground 
motion. Using the GM models~\cite{GM}, 0.2 s of GM is applied to both e$^{-}$ linac+BDS and e$^{+}$ linac+BDS. The resulting luminosity for a single seed of GM as a function of bunch 
number in a train is shown in Fig.~\ref{ILCLuminosity}, where different scenarios of GM are compared. For the noisiest site 
(model C), applying fast position FB stabilisation, a recovery of the luminosity up to 85\% of the nominal value 
is obtained. On the other hand, for quiet sites (models A and B) practically 100\% of the nominal 
luminosity would be achievable with the IP-FB correction. 

\subsection{Results for CLIC}


Latency times of about 20~ns have experimentally been demonstrated by the FONT3 system \cite{FONT3} using an analogue FB processor. Therefore, for the CLIC simulations we have considered a correction iteration every 20~ns. Taking into account the CLIC nominal bunch separation of 0.5~ns, this system performs approximately a correction every 40 bunches ($\approx 8$ iterations per pulse).  

For the CLIC IP position FB simulations, the kicker or corrector has been 
located in the incoming beamline immediately downstream of the final quadrupole QD0. At $\pi/2$ 
phase advance downstream of the IP a BPM, with $1~\mu$m resolution, measures the beam positions. In this case we have employed a FB control loop  based in a simple proportional 
control algorithm. Unlike the angle bunch-to-bunch FB system for ILC, due to latency constraints no angle intra-train FB system has been designed for CLIC.    

\begin{wrapfigure}{r}{0.5\columnwidth}
\centerline{\includegraphics[width=0.55\columnwidth, height=5cm]{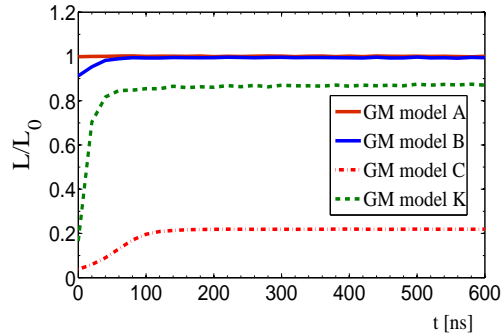}}
\vspace{-0.4cm}
\caption{Relative luminosity versus time for CLIC. Example for a single random seed for different models of ground motion: A, B, C and K.}\label{CLICLuminosity}
\end{wrapfigure}

In this case, applying the model C of GM, simulations of CLIC (Fig.~\ref{CLICLuminosity}) has shown a recovery of about $20\%$ of the nominal luminosity using the intra-train position FB system at the IP. With model K the FB system manages to recover about $80\%$ of the nominal luminosity. For quiet sites (models A and B) practically 100\% of the nominal 
luminosity would be achievable.   

\section{Summary and outlook}

Beam-based feedback systems will be mandatory to achieve the required luminosity goals of the future electron-positron linear colliders. In particular we have simulated the performance of intra-train beam-based FB systems at the IP. These FB systems reduce the transverse beam-beam jitters and steer the electron and positron beams into collision. Simulation results are presented for ILC and CLIC in terms of correcting the transverse vertical jitters generated by ground motion. For quiet sites (models A and B) practically 100\% of the nominal 
luminosity would be achievable using the IP-FB systems. However, assuming the most severe scenario (GM model C), intra-train FB systems at the IP are not enough to achieve the nominal luminosity. This is due to remaining uncorrected pulse-to-pulse jitter, which could in principle be corrected with additional upstream slower FB systems in the BDS. 
  


\section{Acknowledgments}
This work is supported by the Commission of the European Communities under the 6$^{th}$ Framework Programme "Structuring the European
Research Area", contract number RIDS-011899.




\begin{footnotesize}




\begin{thebibliography}{99}

\bibitem{ILC} James Brau (Ed.) {\it et~al.}, {\it ILC Reference Design Report: ILC Global Design Effort and World Wide Study}, arXiv:0712.1959[physics.acc-ph] (2007).
\bibitem{CLIC} Frank Tecker (Ed.) {\it et~al.}, {\it CLIC 2008 Parameters}, CLIC-Note-764 (2008).
\bibitem{guinea} D.~Schulte, PhD Thesis, TESLA-97-08 (1996).
\bibitem{FONT} P.~N.~Burrows {\it et~al.}, {\it Design and Performance of a Prototype Digital Feedback System  for the International Linear Collider Interaction Point}, Proc. of EPAC 2008, Genoa, Italy (2008).
\bibitem{PLACET} D.~Schulte {\it et~al.}, {\it Simulation Package based on Placet}, Proc. of PAC 2001, Chicago, Illinois (2001); \verb$https://savannah.cern.ch/projects/placet$.
\bibitem{THimmel} T.~Himmel, Annu.~Rev.~Nucl.~Part.~Sci.~{\bf 49} 157 (1997).
\bibitem{GM} A.~Seryi {\it et al.}, {\it Simulation Studies of the NLC with the Improved Ground Motion Models}, Proc. of LINAC 2000, Monterey, California (2000).
\bibitem{FONT3} P.~N.~Burrows {\it et~al.}, {\it Tests of the FONT3 Linear Collider Intra-train Beam Feedback System at the ATF}, Proc. of PAC 2005, Knoxville, Tennessee (2005).
\end{thebibliography}
%

\end{footnotesize}


\end{document}